\documentclass[superscriptaddress,twocolumn,printnumbers,amsmath,amssymb,showpacs,10pt]{revtex4}

\newcommand{\ee}{\end{equation}}
\newcommand{\ba}{\begin{eqnarray}}
\newcommand{\ea}{\end{eqnarray}}
\usepackage{slashed}
\usepackage{graphicx}
\usepackage{dcolumn}
\usepackage{bm}
\usepackage{setspace}
\usepackage{color}
\usepackage{etoolbox}
\usepackage{hyperref}

\newcommand{\sech}{\text{sech}}
\newcommand{\Hm}{\mathcal{H}}
\newcommand{\Qm}{\mathcal{Q}}

\begin{document}

\title{Nuts and bolts of supersymmetry}

\author{Nitin Upadhyaya}
\affiliation{James Franck Institute, The University of Chicago, Chicago, Illinois 60637, USA}
\affiliation{Department of Physics, The University of Chicago, Chicago, Illinois 60637, USA}
\affiliation{Center for mathematical modeling, Flame University, Pune, Maharashtra 412115, India}

\author{Bryan G. Chen}
\affiliation{Institute Lorentz for Theoretical physics, Leiden University, Leiden 2333 CA, The Netherlands}

\author{Vincenzo Vitelli}
\email{vitelli@uchicago.edu}
\affiliation{James Franck Institute, The University of Chicago, Chicago, Illinois 60637, USA}
\affiliation{Department of Physics, The University of Chicago, Chicago, Illinois 60637, USA}

\begin{abstract} 

A topological mechanism is a zero elastic-energy deformation of a mechanical structure that is robust against smooth changes in system parameters. Here, we map the nonlinear elasticity of a paradigmatic class of topological mechanisms onto linear fermionic models using a supersymmetric field theory introduced by Witten and Olive. Heuristically, this approach consists of taking the square root of a non-linear Hamiltonian and generalizes the standard procedure of obtaining two copies of Dirac equation from the square root of the linear Klein Gordon equation. Our real space formalism goes beyond topological band theory by incorporating non-linearities and spatial inhomogeneities, such as domain walls, where topological states are typically localized. By viewing the two components of the real fermionic field as site and bond displacements respectively, we determine the relation between the supersymmetry transformations and the Bogomolny-Prasad-Sommerfield (BPS) bound saturated by the mechanism. We show that the mechanical constraint, which enforces a BPS saturated kink into the system, simultaneously precludes an anti-kink. This mechanism breaks the usual kink-antikink symmetry and can be viewed as a manifestation of the underlying supersymmetry being half-broken. 

\end{abstract}
\pacs{45.70.-n, 61.43.Fs, 65.60.+a, 83.80.Fg}
 
\maketitle
\section{Introduction}
Mechanisms are deformations of a mechanical structure which cost zero elastic energy \cite{Thorpe,kaipnas,Maha,Guest,chris}. As an example, consider the folding motion of networks of bars or plates constrained by pivots or hinges around which two adjacent components can freely rotate. When actuated by intrinsic noise or motors and other external fields, such mechanisms could mimic self-propelled motion \cite{Woodhouse_2018} and become the building blocks of robots and smart metamaterials \cite{Fruchart_2018,Rocklin_2018}. Thus, the hard problem of predicting the effect of constraints on an interacting many-body system is as deeply rooted in mechanical design and robotic control theory as it is in modern theoretical physics \cite{Dirac,Faddeev}. 

Here, we study a special class of mechanisms called topological mechanisms which arise through an intriguing correspondence between the bulk and the boundary (or defects) of {\it periodic} mechanical structures on the verge of stability \cite{kanelubensky}. Such mechanisms are robust to smooth changes in material parameters so long as the global connectivity of the structure is preserved \cite{vincenzo,kanelubensky,Huber_2016,Katia_2017,Bryan,Paulose,Yujie}. Inspired by the study of electronic topological materials \cite{Ashwin,Theo,vladimir,kanecolloquium}, topological mechanical states are now being engineered that not only display many of the features originally thought to be exclusively in the domain of quantum condensed matter, but also provide novel ideas and phenomena, often using easy to assemble components such as Lego blocks \cite{Mousavi2015,Susstrunk2016,Chen2016,Fruchart_2020,Deymier2015,Bi2015,Peano2015,Wang2015a,Wang2015b,Po2014,Meeussen2016,Chen2016,Susstrunk2015,Khanikaev2015,Rocklin2015,Kariyado2015,Rocklin2015a,Nash2015,Mao_2018,Fruchart_2020,Guido_2018,Bartolo_2019}.

Unlike their electronic counterparts, topological mechanisms are not adequately addressed by the theory of linear vibrations for the following reasons. First, mechanisms often involve large deformations of the mechanical structure and hence nonlinearities become paramount. While Maxwell's constraint counting theory applied to the linear vibrational spectrum can predict the presence of mechanisms, it does not describe their properties \cite{Stephan}. Second, topological mechanisms can exist in structures that are not periodic \cite{Rocklin2015a}. For example, domain walls or kinks that cost zero stretching energy can propagate through a one dimensional topological \cite{Bryan} chain even in the presence of strong disorder \cite{Yujie}, provided that no bonds are broken or states of self-stress created.

In this article, we propose a ``classically relevant'' supersymmetric (SUSY) extension of nonlinear continuum mechanics that allows one to a-priori keep track of internal degrees of freedom and deformations that inevitably accompany the dynamics of extended classical excitations, such as kinks and solitons. We illustrate this idea using the quasi one-dimensional topological mechanism as a paradigmatic example of a classical system whose kink solution saturates the Bogomolny-Prasad-Sommerfield (BPS) bound \cite{Bogo,Vachaspati,ana}. Our approach consists in mapping the one component boson (described by a nonlinear Klein-Gordon theory) to a two component Majorana field (Dirac equation with real solution) via Dirac's square root procedure. We identify the square root of the Hamiltonian with one of the conserved charges in the Witten-Olive supersymmetric (SUSY) field theory -- a proposed spacetime symmetry which relates bosons and fermions \cite{witten0,witten1,witten2,tarun,Lawler_1,Lawler_2}, and in the process, obtain another conserved charge, which we associate with a partner Hamiltonian. In the BPS saturated case, only one of these charges is zero and therefore, the supersymmetry is half-broken. Further, we show that the two components of the Majorana field physically correspond to the kink-induced displacement and stress fields respectively. These fields in turn are supersymmetric partner modes in an underlying (fluctuations around the kink field) quantum mechanical supersymmetry, where we again find that supersymmetry is half-broken. We identify this hierarchical breaking of supersymmetry with the BPS bound being saturated due to the breaking of space inversion symmetry in the underlying lattice and the resulting large energy gap between kink and antikink and correspondingly, between displacements and stresses.

\section{Linear topological mechanics} The first step towards studying mechanisms in an arbitrary mechanical structure is to identify (within linear theory) the zero energy eigenvalues (modes) of the Fourier transformed rigidity matrix (or equivalenty dynamical matrix, see Appendix) which within linear order, relates bond stretching to site displacements \cite{demaine}. Physically, a zero mode causes no stretching of the elastic bonds even when some of the sites are displaced. Conversely, a state of self-stress is an assignment of bond tensions that does not result in site displacements. The generalized Maxwell-Calladine relation \cite{Calladine} stipulates that for $N$ sites in $d$-dimensions, the number of zero modes, $n_\text{m}$ minus states of self-stress, $n_\text{ss}$ equals the number of degrees of freedom $N_\text{df}=dN$ minus constraints $N_\text{c}$ 

\begin{equation}
\nu \equiv n_{m}- n_{ss}=N_\text{df} - N_\text{c}. 
\label{eq1}
\end{equation}

A structure is rigid, floppy or isostatic, depending on whether $\nu<0$, $\nu>0$ or $\nu=0$ respectively \cite{Thorpe,Andrea}.
Once the connectivity is fixed (which fixes the right-hand side of Eq.\ (\ref{eq1})), the index $\nu$ can be viewed as a topological charge \cite{kamien}, invariant under smooth deformations of the local bond length. Kane and Lubensky laid the foundations for the use of topological band theory to establish the topological nature and location of zero modes associated with so called Maxwell lattices \cite{kanelubensky}. While the study of mechanical topological modes began with phonons and hence, linear vibrations in mind, zero modes could be either infinitesimal or finite motions of the structure and only a nonlinear analysis can reveal their full nature.


\section{Non-linear mechanics of topological chains} 
\begin{figure}
\includegraphics[width=.5\textwidth]{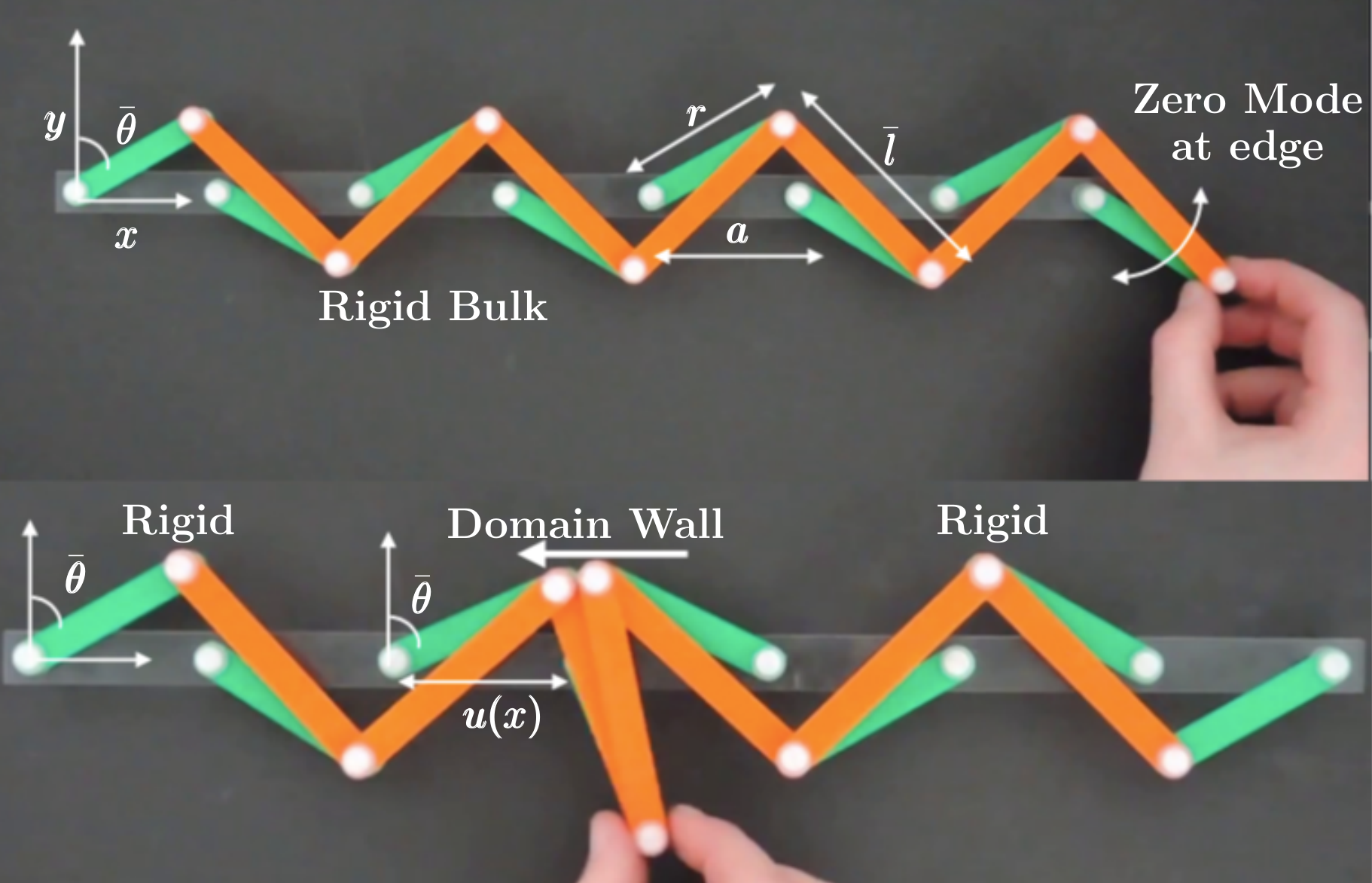} 
\caption{\label{Kink}A mechanism inspired by the organic molecule poly- acetylene is constructed from rigid (green) rotors coupled by (orange) bars (see Movies in SI). Once actuated (by hand here), the zero-energy mode travels down the chain (indicated by arrows). A domain wall separates the left and right leaning green bars. Here, $r$ is the length of the rotor, $a$ is the lattice spacing, $\theta(x)$ is the angle that the rotor at position $x$ makes with respect to the vertical and $u(x)=r\sin\theta(x)$ is the projection of the rotor length along the $x-$ axis.}
\end{figure}

Before discussing the supersymmetric field theory, we briefly review the relation between zero energy modes and nonlinear mechanisms for a quasi one-dimensional mechanism, see Fig.\ (\ref{Kink}) \cite{kanelubensky,Bryan}. This realization consists of a periodically repeating, dimerized unit of green rigid rotors (alternately pointing up and down), each of length $r$ separated by a distance $a$ and constrained to rotate about fixed white bolts. The local orientation of each rotor with respect to the vertical is denoted by $\theta(x)$ and their horizontal projection by $u(x)=r \sin{\theta(x)}$. The rigid rotors are coupled to their nearest neighbors through orange bars that can be viewed as Hookean springs with elastic constant $k\rightarrow \infty$. There is exactly one fewer constraint than degrees of freedom in the chain, thus by Eq.\ (\ref{eq1}) there is exactly one zero-energy mode \cite{footnote2}. If we denote by $\bar{\theta}$ the angle that the rotors make with the vertical in the initial uniform configuration (assumed positive in the clockwise sense), then the zero energy mode will be localized to the left edge if $\bar{\theta}<0$ or right edge if $\bar{\theta}>0$.

In order to derive the continuum theory, we express the length $l$ of the rigid bar that connects two adjacent rotors in terms of $r,a$ and their respective angular displacements and solve for the rigid bar constraint, i.e., $l=\bar{l}$, where $\bar{l}$ is the equilibrium length of the orange bars in the uniform state where $\theta(x)=\bar{\theta},\pi-\bar{\theta}$, see Fig.\ (\ref{Kink}) and Appendix B. In the limit that $2r\sin\bar{\theta}\ll a$ and $a\ll1$, we find the following nonlinear differential equation for $u(x)=r\sin\theta(x)$ \cite{Bryan}:
\begin{align}
\frac{a^2}{2}\frac{du}{dx}=u^2 - \bar{u}^2,
\label{constraint}
\end{align}
where $\bar{u}=r\sin\bar{\theta}$. The solution of this nonlinear differential equation (up to a constant) is a kink (domain wall) 
\begin{align}
u_s = -\bar{u}\tanh\left(\frac{x-x_0}{\frac{a^2}{2\bar{u}}}\right), \label{static_kink}
\end{align}
which interpolates between the two topologically distinct uniform states of the chain $u(x\rightarrow\infty)=-\bar{u}$ and $u(x\rightarrow -\infty)=\bar{u}$. Moreover, the kink can translate along the chain by a sequential activation of the joints, see Fig.\ (\ref{Kink}). The dynamics is described by relaxing the rigid bar constraint and introducing a finite spring constant $k$ (for orange bars) to obtain the nonlinear Hamiltonian \cite{Bryan} 
\begin{align}
\mathcal{H} = \frac{H}{\rho} = \frac{1}{2}\int dx \left[\pi^2 + c^2\left(\frac{\partial u}{\partial x} +\sqrt{V(u)} \right)^2 \right]
\label{eqw}
\end{align}
where, we have re-scaled the original Hamiltonian by the mass density $\rho=\frac{M}{a}$ ($M$ being the mass of the rotors) and defined the conjugate momentum field $\pi(x,t)=\frac{\partial u}{\partial t}$, linear sound speed $c=\frac{a^2}{\bar{l}}\sqrt{\frac{k}{M}}$ and
\begin{align}
V(u)= \frac{4}{a^4}(u^2 - \bar{u}^2)^2.
\label{dw}
\end{align}
Note, that the Hamiltonian in Eq.\ (\ref{eqw}) is the sum of two perfect squares. Consequently, the static kink configuration in Eq.\ (\ref{static_kink}) that solves the first order constraint equation \ (\ref{constraint}) can be simply obtained by setting to zero the term within round brackets in Eq.\ (\ref{eqw}). 

The approach we adopt was first proposed by Bogomolny \cite{Bogo,Vachaspati}. This method that we briefly review is used to directly obtain first order equations that yield static kink (and antikink) configurations of the field (without finding the equations of motion). It is instructive to compare
Eq. (\ref{eqw}) to an Hamiltonian of the form 
\begin{eqnarray}
\mathcal{H} = \frac{1}{2}\int dx \left[\pi^2+\left(\frac{\partial u}{\partial x}\right)^2 + V(u)\right],
\label{normal_phi4}
\end{eqnarray}
If a double well potential of the form in Eq. (\ref{dw})
is chosen, Eq. (\ref{normal_phi4}) describes an Ising model. Upon completing the square one obtains
\begin{eqnarray}
\mathcal{H} = \frac{1}{2}\int dx \left[\pi^2+\left(\frac{\partial u}{\partial x}\mp \sqrt{V(u)}\right)^2\right]\pm \nonumber \\ \int du \sqrt{V(u)} \!\!\!\! \quad .
\label{boundary}
\end{eqnarray}
Eq. (\ref{boundary}) reduces to Eq. (\ref{eqw}) (for $c=1$) aside from the last term in Eq. (\ref{boundary}) which, as we shall see, is a boundary term. 
If $u(x\rightarrow\infty)=\pm\bar{u}$, then energy is minimized if 
\begin{align}
\pi=0, \quad \frac{\partial u}{\partial x}\mp \sqrt{V(u)}=0. \label{symm}
\end{align}
The minimum energy $E$ of the corresponding field configuration is then given by
\begin{align}
E=\pm\int du \sqrt{V(u)}, \label{symm2}
\end{align}
where the plus and minus signs correspond to {\it static}
kink and antikink solutions respectively. It is convenient to define a function $W(u)$ that satisfies
\begin{align}
\left(\frac{\delta W}{\delta u}\right)^2 \equiv V(u) .
\label{V}
\end{align}
With this identification, the {\it minimum} energy $E$ in Eq. (\ref{symm2}) is equal to a topological charge $Z$ defined by
\begin{eqnarray}
Z &\equiv& \int^{\infty} _{-\infty} du \left(\frac{\delta W}{\delta u}\right)  \\
   &=&  W(u(\infty))-W(u(-\infty)).\label{Topological_Charge}
\end{eqnarray}
The quantity $Z$ is topological in the sense that it depends purely on the boundary values of the field $u(x)$ at $x=\pm \infty$ and not on the field profile. Thus, the energy (or mass) of a {\it dynamic} kink(antikink) must be greater or equal (in the quasistatic limit) to $Z$. This is known as the Bogomolny-Prasad-Sommerfield (BPS) bound \cite{Bogo,Vachaspati,ana}. Note that the two first order field solutions Eq.\ (\ref{symm}) and the associated field energies are symmetric between a kink ($+$) and an antikink ($-$). 

In the case of Eq.\ (\ref{eqw}), the kink solution (but not the antikink) makes the elastic energy {\it vanish} and there is no additional boundary term. The kink is said to saturate the BPS bound. At the same time, the apparent symmetry between a kink and anti-kink ($u\rightarrow -u$) in Eq.\ (\ref{eom_2}) no longer exists, since the elastic energy in Eq.\ (\ref{eqw}) only vanishes for the specific (static) configuration which satisfies Eq.\ (\ref{constraint}), while it costs a finite energy for the other. Physically, a kink state in the bulk of the chain corresponds to right leaning bars (green) on the left half of the chain and left leaning bars, on the right side, with a nearly vertical bar in the middle of the domain wall, see Fig.\ (\ref{Kink}). However, an anti-kink state will require left leaning bars on the left side of the chain and right leaning bars on the right side with a nearly vertical bar in the middle, and will thus require the orange connecting bars to be either, of longer rest length or, be stretched, see Ref. \cite{Bryan,Yujie} for more details and pictures. Since the kink profile is the finite amplitude manifestation (in nonlinear theory) of the zero energy edge mode (within linear theory), it can propagate down the chain without costing any energy. Thus, the asymmetry between the kink and anti-kink can be ultimately traced back to the existence of the localized edge mode under open boundary conditions, consistent with Eq.\ (\ref{eq1}) and the breaking of spatial symmetry by the underlying lattice. 

\begin{figure}
\includegraphics[width=.45\textwidth]{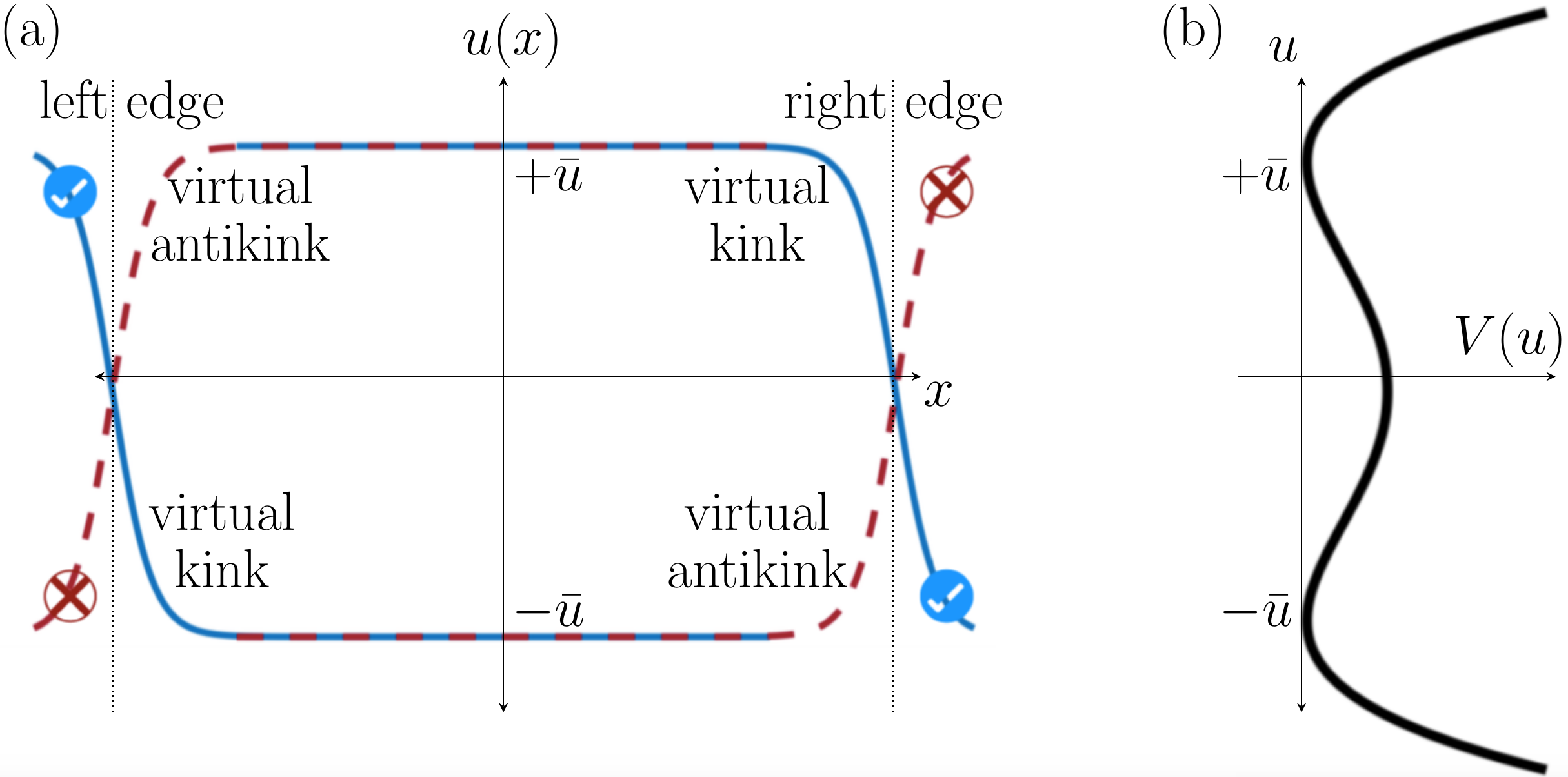} \\
\par\vspace{0.5cm}
\includegraphics[width=.45\textwidth]{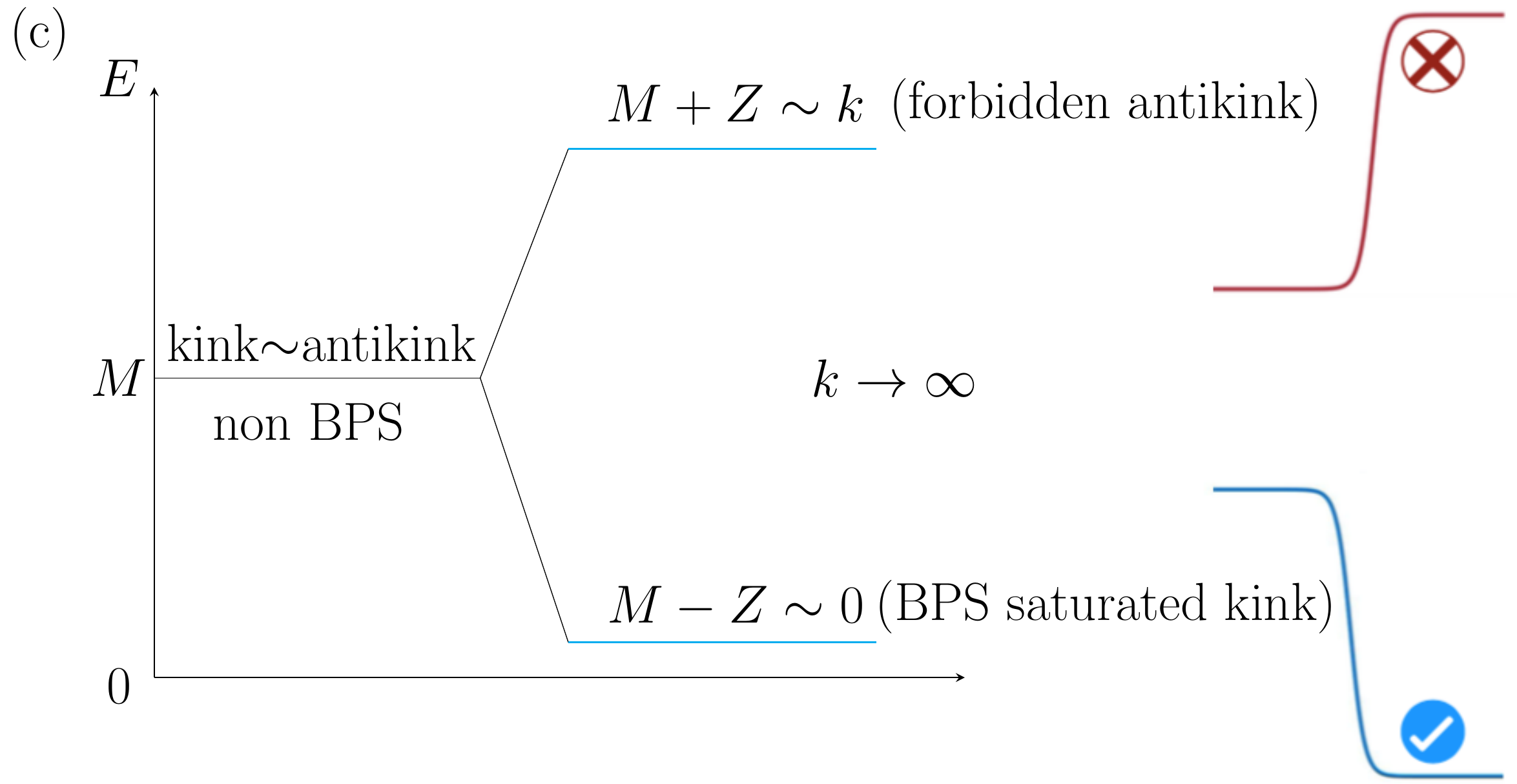} 
\caption{\label{Energy}(top):(a) Illustrating the asymmetry between allowed kink states (solid curves) and forbidden anti-kink (dotted curves) states. A zero mode (virtual kink) localized at the right edge (uniform potential $+\bar{u}$ in the bulk) can propagate down the chain as a kink, and localize at the left edge (uniform potential $-\bar{u}$ in the bulk)(b) The double well potential corresponding to the two kink-states. (bottom): Energy diagram illustrating the kink-anti-kink asymmetry. As the spring constant $k\rightarrow\infty$, it takes an infinite energy to excite an anti-kink state.}
\end{figure}

We illustrate this crucial symmetry breaking in Fig.\ (\ref{Energy}). In the top panel, we show the allowed and forbidden zero mode configurations in the chain. According to the linearized theory, for a configuration with $\bar{u}>0(<0)$, the zero mode is initially ``localized'' at the right(left) edge. In Fig.\ (\ref{Energy}), we refer to these edge-localized configurations as {\it virtual} kink and anti-kink respectively since in this state, only a part of their full profile is visible. However, when the nonlinear nature of the mechanical structure is taken into account, we find the zero mode develops into a {\it real} kink that can propagate down the chain and transition between the right and left localized states (or manifest as an intermediate state in the bulk of the chain, see also Fig.\ (\ref{Kink})). However, we never find a configuration which supports an anti-kink. This asymmetry is physically the result of a finite energy gap between the kink and anti-kink configurations, see Fig.\ (\ref{Energy}) (bottom panel). In the next sections we demonstrate that, in the supersymmetric version of the field theory, this asymmetry between kink and anti-kinks is related to a breaking of the supersymmetry.

The Euler-Lagrange equation of motion for Eq.\ (\ref{eqw}) does not depend on the boundary term and yields the nonlinear Klein-Gordon equation
\begin{eqnarray}
\frac{\partial^2 {u}}{\partial t^2}- c^2\frac{\partial^2 u}{\partial x^2} = -\frac{8c^2}{a^4}u(u^2-\bar{u}^2) \label{eom_2}
\end{eqnarray}
The domain wall solution to Eq.\ (\ref{eom_2}) interpolates between left-leaning and right-leaning rotors, see Fig.\ (\ref{Kink}) and carries with it the zero energy mode as it propagates down the chain. This is reminiscent of how a domain wall facilitates electron transport in poly-acetylene \cite{ssh}. In poly-acetylene, however, the kink is associated with bond distortions.

\begin{figure}
\includegraphics[width=.45\textwidth]{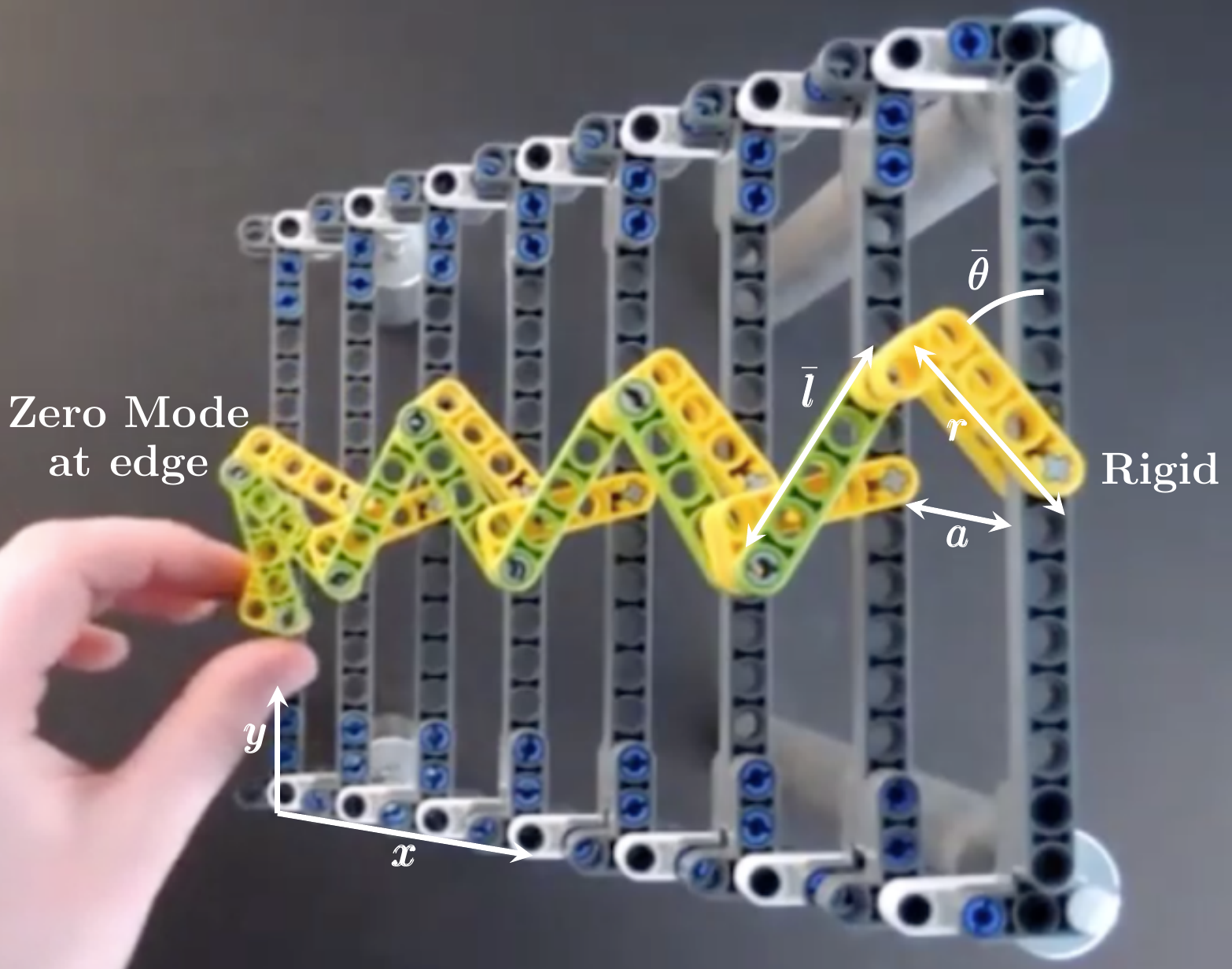} 
\caption{\label{Spinor} A Lego realization of a structure where a zero energy edge mode evolves into a Sine-Gordon soliton that can start only from one of the two boundaries of the system \cite{Bryan}. During the propagation of this soliton, the rotating bars (yellow) undergo a rotation by $\pi$ and thus requires a three dimensional realization. The soliton is obtained as the solution of the constraint equation Eq. (\ref{SG_constraint}). Here, $\bar{\theta}$ is the equilibrium angle that the yellow rotors make with the vertical gray bars, $a$ is lattice spacing along $x-$direction, $r$ is the length of the yellow bars and the length of green connecting bars is $\bar{l}=(a^2+4r^2\cos^2\bar{\theta})^{\frac{1}{2}}$, see Appendix B. These Sine-Gordon kinks are obtained in the limit $r\gg a$. For a video, see SI.}
\end{figure}

While we have taken the Ising-like kink as our main working example, we can apply the formalism developed in this paper also to other structures \cite{Bryan} like the helical realization in Fig.\ (\ref{Spinor}). In contrast to Fig.\ (\ref{Kink}), this structure realizes the opposite limit $r \gg a$, and here we obtain the following non-linear constraint equation 
\begin{align}
\frac{d\theta}{dx}=-\frac{1}{r}\frac{\sin(\theta-\bar{\theta})}{\sin\bar{\theta}}, \label{SG_constraint}
\end{align}
instead of Eq.\ (\ref{constraint}). The resulting dynamics is governed by the Sine-Gordon equation \cite{Bryan}. In contrast with the Ising-like kink discussed so far, we now have spinning solitons: the rotating bars (shown in yellow in Fig.\ (\ref{Spinor})) undergo a rotation by $\pi$ and thus the dynamics is described in terms of the full angle $\theta (x)$ which the rotating bars make with the vertical. As seen in Fig. \ (\ref{Spinor}), in order to allow the rotating bars to rotate by $\pi$ without getting obstructed by the rigid base (contrast with Fig.\ (\ref{Kink})), we need a staggered arrangement of bars, effectively embedding the structure in three dimensions. Rotating the zero energy edge mode (shown in the far left end in Fig.\ (\ref{Spinor})) by $\pi$ shifts the zero mode to the adjacent rotor (along the $x-$ axis), whose dynamics can effectively still be approximated by two copies of the one dimensional Sine-Gordon equation, one copy each for the rotors on odd and even sites respectively, see Ref.\cite{Bryan} for more details. For each copy of the Sine-Gordon soliton, the supersymmetric extension discussed in the next section carries through by replacing $u(x)\rightarrow \theta(x)$, with 
\begin{align}
\frac{\delta W(\theta)}{\delta\theta}=\pm\frac{1}{r}\frac{\sin(\theta\mp\bar{\theta})}{\sin\bar{\theta}}
\end{align}
in Eqs. (\ref{V}) and \ (\ref{eqw}) for the odd (+) and even (-) sites. 

\section{Supersymmetric field theory}
In order to further develop the connections between BPS kinks and supersymmetry and the emergence of fermionic variables, we apply Dirac's procedure to take the square root of the classical Hamiltonian in Eq.(\ref{eqw}). Note that, taking the square root of an equation of motion (where possible) is a useful method to simplify and reduce the order of an equation. For instance, the square root of the linear wave equation gives two first order equations which are then used to construct the d'Alembert's solutions. Likewise, the BPS method discussed in the last section allows us to obtain first order equations directly from the Hamiltonian. Moreover, Eq.\ (\ref{Topological_Charge}) indicates that for a kink configuration, the topological charge is defined via the square root of the potential term. Taking inspiration from these, we find that applying Dirac's square root procedure to a general Hamiltonian results in a dynamical charge if we allow anti-commutating variables in the theory. As we discuss next, this  charge corresponds to one of the conserved charges in a supersymmetric extension of the original classical theory. For a SUSY theory with two fermionic variables, we obtain as a byproduct, a second conserved charge, which defines a partner Hamiltonian to the original.

Note, the existence of a BPS kink (Eq.\ (\ref{static_kink})) has allowed us to express the elastic term in Eq.\ (\ref{eqw}) as a perfect square. Thus, we define a field theoretic charge $Q_1$ of the form 
\begin{align}
Q_1= \int \ dx \left[\pi \psi _1 + c\left(\frac{\partial u}{\partial x}+ \frac{\delta W}{\delta u}\right)\psi _2\right], \label{charge_field}
\end{align}
where, we introduce two real field variables $\psi _{1,2}(x,t)$ and a potential $W(u)$ which equals $W(u) = \frac{2}{a^2}(\bar{u}^2 u - \frac{1}{3} u^3)$. We refer the reader unfamiliar with supersymmetry to the pedagogical treatment in Ref. \cite{Shifman} whose approach and notation we follow closely. In Appendix A, we show that in order for $Q_1^2=\mathcal{H}$, $\psi _{1,2}$ needs to satisfy the equal time anti-commutation relations:
\begin{align}
\left\{\psi _{a}(x,t),\bar{\psi} _{b}(x',t)\right\}=\left(\gamma^0\right)_{ab}\delta(x-x'), \label{anticommute}
\end{align}
where the index $a,b=\{1,2\}$ and $\bar{\psi}_1 = i\psi _2$ and $\bar{\psi}_2=-i\psi _1$. Here and in the following, we make use of the following gamma matrices:
\begin{align}
\gamma^0=\sigma_2=\begin{pmatrix}
0  & -i \\
i   & 0 \\
\end{pmatrix}, \quad
\gamma^1=i\sigma_3=\begin{pmatrix}
i & 0\\
0 & -i\\
\end{pmatrix}.
\end{align}
Moreover, we can combine $\psi _{1,2}\equiv\psi _{1,2}(x,t)$ into a two component Majorana field $\Psi\equiv \Psi(x,t)=
\begin{pmatrix}
\psi_{1} \\
\psi_{2} \\
\end{pmatrix}$
with its conjugate defined as $\bar{\Psi}\equiv \Psi^{\dagger}\gamma ^0$. Note again, $\psi _{1,2}$ are real and therefore, we refer to $\Psi$ as a Majorana field (the particle is the same as the antiparticle) \cite{Shifman}. 

 

Consider next the supersymmetric Lagrangian \cite{witten2}- 
\begin{align}
\mathcal{L}_s = \mathcal{L}_b + \mathcal{L}_f  \label{lagrangian},
\end{align} 
where $\mathcal{L}_b$ is the bosonic part of the Lagrangian obtained from Eq.\ (\ref{eqw})
\begin{align}
\mathcal{L}_b= \frac{1}{2}\int dx \left[\left(\frac{\partial u}{\partial t}\right)^2 - c^2\left(\frac{\partial u}{\partial x}-\frac{\delta W}{\delta u} \right)^2 \right].
\end{align}
while the Lagrangian $\mathcal{L}_f$ is expressed in terms of the Majorana field $\Psi$
\begin{align}
\mathcal{L}_f = \frac{1}{2} \int dx \ i \bar{\Psi} \gamma^{\nu} \partial_{\nu} \Psi-\frac{\delta^2 W}{\delta u^2} \bar{\Psi}{\Psi},
\end{align}
where $\nu=\{0,1\}$ denote time ($t$) and space ($x$) components respectively, $\partial _{0}\rightarrow\partial _t$ and $\partial _1\rightarrow c\frac{\partial }{\partial x}$ and $\bar{\Psi}=\Psi^{\dagger}\gamma^0$.

The action of Eq.\ (\ref{lagrangian}) is invariant under supersymmetry transformations \cite{Augirre_2018}
\begin{align}
\delta u &=& i\epsilon _2\psi _1 - i\epsilon _1\psi _2 , \\
\delta\psi_1 &=& -\dot{u}\epsilon _2 + \epsilon _1\left(u' - \frac{\delta W}{\delta u}\right), \\
\delta\psi_2 &=&  \dot{u}\epsilon _1 - \epsilon _2\left(u'  + \frac{\delta W}{\delta u}\right),  \label{SUSY_transform}
\end{align}
where $\epsilon _{1,2}$ are real anti-commutating transformation parameters. The two conserved charges associated with this supersymmetry are $Q_1$ (Eq.\ \ref{charge_field}) and 
\begin{align}
Q_{2} &=& \int \ dx \left[\pi\psi _2 + c\left(\frac{\partial u}{\partial x} - \frac{\delta W}{\delta u}\right)\psi _1\right]. \label{charge_field2}
\end{align}
Note, $Q_1$ and $Q_2$ have different signs of the potential term, i.e., $\pm\left(\frac{\delta W}{\delta u}\right)$ respectively. Thus, they square to generate two different Hamiltonians - 
\begin{align}
\mathcal{H}_{1,2}=\frac{1}{2}\int dx \ \pi^2+c^2\left(\frac{\partial u}{\partial x}\right)^2+c^2\left(\frac{\delta W}{\delta u}\right)^2 \pm\nonumber \\ 2c^2\frac{\partial u}{\partial x}\frac{\delta W}{\delta u}.
\label{Q2}
\end{align}
While $\mathcal{H}_1$ is the same as Eq.\ (\ref{eqw}), $\mathcal{H}_2$ corresponds to a Hamiltonian generated from a constraint which yields an anti-kink profile, i.e., Eq.\ (\ref{static_kink}) with $u\rightarrow -u$. In supersymmetric theories $\mathcal{H}_{1,2}$ are called partner Hamiltonians, see Fig.\ (\ref{Schematic}). Note, how the first three terms in Eqs.\ (\ref{Q2}) reproduce the well known ``$\phi ^4$'' theory (eg: in the SSH model and discussed in Eq.\ (\ref{normal_phi4})) whose integral is the energy (or mass) of the kink or anti-kink solution which we denote by $M$ \cite{Shifman}. The last term in both the equations however is special since it is a total derivative whose integral only depends upon the boundary conditions and hence it is a topological property of the mechanical structure, see Eq.\ (\ref{Topological_Charge}). Thus, $Q^2_1=M+Z$ and $Q^2_2=M-Z$.
Without the kinetic term (static solutions) in Eq.\ (\ref{eqw}), $M\rightarrow M_s$ in which case $E=M_s\pm Z$ represents the elastic potential energy associated with the (anti)kink configurations. In the special case when either $M=Z$ or $M=-Z$, the elastic potential energy associated with one of the configurations is zero and this is the defining feature of a mechanism where the (BPS) bound ($M\geq Z$) is saturated. In our case, the condition $M=-Z$ recovers the constraint equation Eq.\ (\ref{constraint}) which we refer to as the kink profile. The anti-kink profile then has an elastic energy $E=2Z$ and therefore, the kink and anti-kink configurations are not symmetrical. In our framework, the kink-antikink are obtained from partner Hamiltonians defined through a supersymmetric extension of the classic theory. 

Note also from Eq.\ (\ref{SUSY_transform}), the variation of the $\Psi$ field vanishes only if the BPS equation is satisfied for a static kink. In other words, the classical BPS solutions remain invariant under supersymmetry transformations, and thus supersymmetry is said to be half broken.



 \section{ Supersymmetric quantum mechanics} 
 In order to further clarify supersymmetry breaking (due to BPS saturated kink) and the physical meaning of the Majorana field in the mechanical context, we next study small fluctuations around the kink. In the process, we reveal another supersymmetric structure inherent to the study of fluctuations, i.e., at the particle level, referred to as supersymmetric quantum mechanics. 
 
As the first step, we linearize Eq.\ (\ref{eom_2}) around the kink solution $u _s(x,t)$ by expressing $u(x,t) =u _s(x,t) + \psi(x,t)$ and look for small distortions of the kink field in the form $\psi(x,t)= \psi^{(1)}_n(x)\text{exp}(i\omega_n t)$. This in turn yields a Schr\"odinger-like equation for $\psi(x,t)$
\begin{eqnarray}
\mathcal{H}_1\psi^{(1)}_n = \rho \omega_n^2\psi^{(1)}_n \label{second_order}
\end{eqnarray}
where, $\mathcal{H}_1=c^2\left(-\frac{\partial ^2}{\partial x^2} + U_1\right)$ is the second-order differential operator and we have defined the potential $U_1 =  \left[\left(\frac{d\tilde{V}}{du}\right)^2 + \tilde{V}\frac{d^2\tilde{V}}{du^2}\right]_{u=u_s(x)}$, with, $\tilde{V}(u)=\frac{\delta W}{\delta u}$. Note, since $\mathcal{H}_1$ is a Hermitian differential operator, its eigenvectors $\psi^{(1)}_n$ constitute an orthogonal basis. In particular, the bound state solutions $\psi^{(1)}_b$ are real and satisfy the orthogonality condition $\psi^{(1)} _b(x)\psi^{(1)} _b(x') \propto \delta (x-x')$. 

\begin{figure}
\includegraphics[width=.5\textwidth]{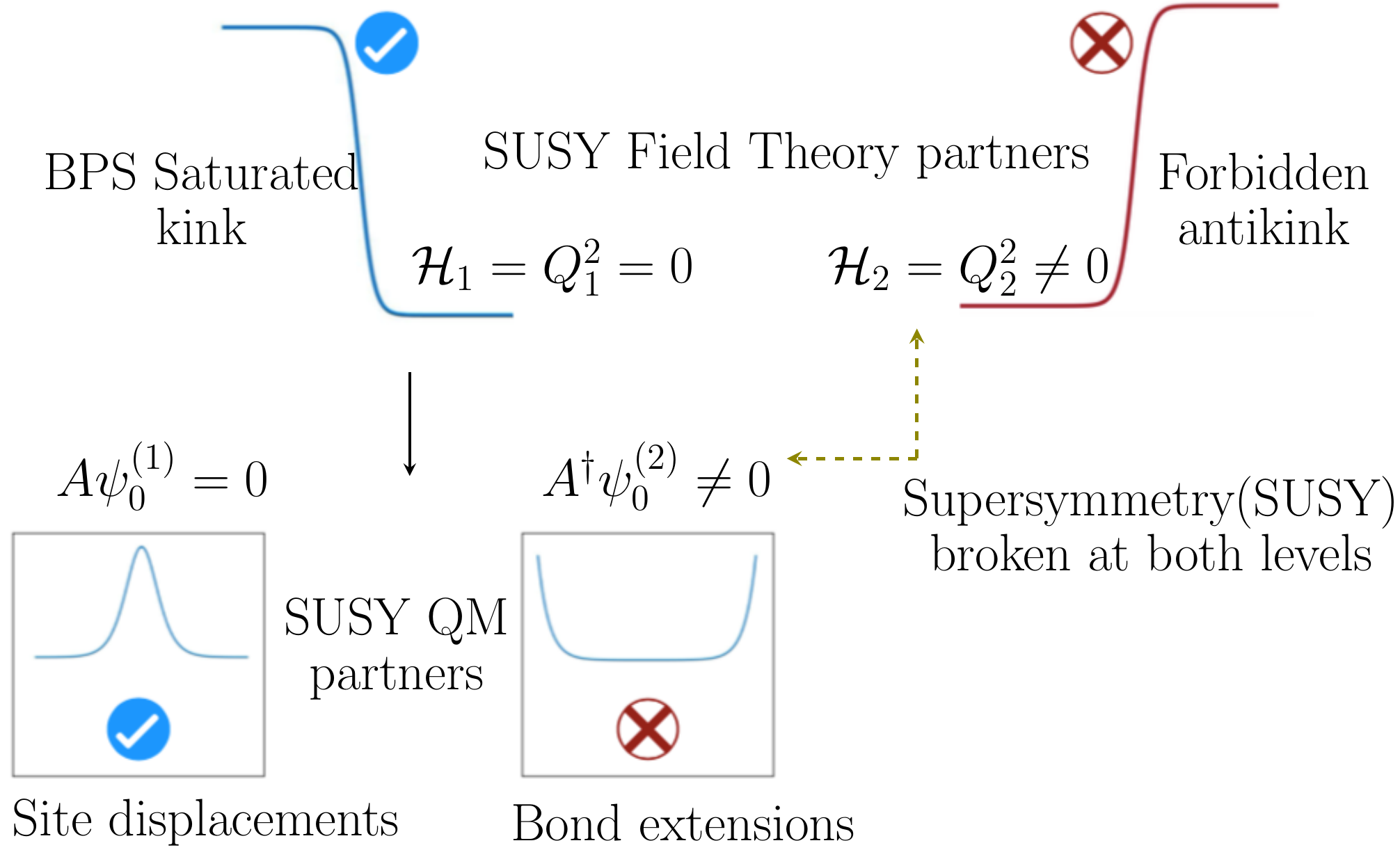} \\
\caption{\label{Schematic} Illustrating the half broken supersymmetry at both the field theoretic level (associated with the kink- antikink asymmetry) and in describing the fluctuations around the classical kink (associated with absence of bond stretching in mechanisms). }
\end{figure}

Next, by defining $w(x)=-\frac{\delta^2W}{\delta u^2} \label{w}$, we can factorize $\mathcal{H}_1$ as a product of two first-order differential operators, ${\cal H}_1=A^{\dagger}A$, where
\begin{eqnarray}
A = -\frac{d}{dx} + w(x), \quad
A^{\dagger}= \frac{d}{dx}+w(x). \label{A_operator}
\end{eqnarray}
Note, Eqs. \ (\ref{A_operator}) have a structure very similar to the static part of the charges $Q_{1,2}$ Eq. (\ref{charge_field}). Since $A^{\dagger}A\neq AA^{\dagger}$, we can define a Hamiltonian $\mathcal{H}_2 =  AA^{\dagger} $  with potential  $U_2(x) = \left[\left(\frac{d\tilde{V}}{du}\right)^2 - \tilde{V}\frac{d^2\tilde{V}}{du^2}\right]$. Together $\mathcal{H}_{1,2}$ constitute a pair of quantum supersymmetric partner Hamiltonians, see also Fig.\ (\ref{Schematic}). In analogy with the notation used in quantum mechanics, we label operator $A$ as a lowering operator and $A^{\dagger}$ as a raising operator and discuss next their physical meaning for our mechanical system.


Physically, the bound states of $\mathcal{H}_1(x)$ are the site displacement eigenfunctions. Applying the lowering operator once, we obtain the corresponding bond extensions, $\psi^{(2)}_n(x)=A \psi^{(1)}_n$, or equivalently the tensions $k \psi^{(2)}_n(x)$ that would be measured in an elastic structure. The operator $A$ is thus a continuum version of the discrete rigidity matrix $R$, see Appendix B for further details. Alternately, the bond extensions $\psi^{(2)}_n(x)$ are obtained directly as the bound states of $U_2$, the potential associated with the partner Hamiltonian $\mathcal{H}_2$. The Hamiltonians ${\cal H}_1,{\cal H}_2$ are said to be partners because once the eigenvalues (eigenfunctions) of ${\cal H}_1$ are known, the corresponding ones for ${\cal H}_2$ can be easily obtained (except for the zero-energy eigenvalue that we assume to be part of ${\cal H}_1$ as discussed below). Thus for example, if ${\cal H}_1\psi^{(1)}_n=E^{(1)}_n\psi^{(1)}_n$, then ${\cal H}_2(A\psi^{(1)}_n)=AA^{\dagger}A\psi^{(1)}_n=E^{(1)}_n(A\psi^{(1)}_n)$. Each eigenfunction in ${\cal H}_1$ has a partner in the spectrum of ${\cal H}_2$ except for the ground state defined via ${\cal H}_1\psi_0=A^{\dagger}A\psi_0=0$. The site-displacement field is obtained from the bond-stretching field by applying the raising operator, i.e.\ $\psi^{(1)}_n(x)=A^{\dagger} \psi^{(2)}_n$. 

The zero energy bound states which are obtained by solving the pair of Eqs.\ (\ref{A_operator}) with $w(x) = \frac{4u_s}{a^2} = \frac{-4\bar{u}}{a^2}\tanh\left(\frac{2\bar{u}x}{a^2}\right)$ are however special. Solving for $A\psi^{(1)}_0=0$ (equivalently $\mathcal{H}_1\psi^{(1)}_0=0$), we obtain
\begin{align}
\psi^{(1)}_0(x) \propto \sech^2\left(\frac{2\bar{u}}{a^2}x\right)   \label{sech}
\end{align}
while from $A^{\dagger}\psi^{(2)}_0=0$ (equivalently $\mathcal{H}_2\psi^{(2)}_0=0$), we obtain
\begin{align}
\psi^{(2)}_0(x) \propto \cosh^2\left(\frac{2\bar{u}}{a^2}x\right). \label{cosh}
\end{align}
The proportionality constant for Eq.\ (\ref{sech})-(\ref{cosh}) are obtained by requiring the solutions to be normalizable. While Eq.\ (\ref{sech}) is always normalizable, Eq.\ (\ref{cosh}) grows exponentially with the system size and is thus, physically not observable in a large sample (it is localized to the sample edges). In the quantum regime, this ``unpairing'' of the zero modes results in the curios phenomena of fractional fermion number \cite{ssh,JR,JS}. However,  this is to be contrasted with our mechanical system where even if the system size remains small,  Eq.\ (\ref{cosh}) is not physically realizable when we consider mechanisms, i.e., when the spring constant $k \rightarrow \infty$. This is because $A^{\dagger}$ corresponds to bond extensions which are forbidden in the mechanisms limit and thus the only permissible solution to Eq.\ (\ref{cosh}) is
\begin{align}
\psi^{(2)}_0(x)=0. \label{zero}
\end{align}
On the other hand, the orthogonality of the modes leads to a normalization factor $C=\left(\frac{3a^2}{8\bar{u}^3}\right)^{\frac{1}{2}}$ for $\psi^{(1)} _0$ via the integral:
\begin{align}
C^2\left(\frac{2\bar{u}}{a^2}\right)^2\int^{\infty} _{-\infty} \ dx \ \sech^4\left(\frac{2\bar{u}}{a^2}x\right) = 1.
\end{align}

We thus see that only one of $\mathcal{H}_{1}\psi^{(1)}_{0}$ or $\mathcal{H}_{2}\psi^{(2)}_{0}$ can be zero, reminiscent of the asymmetry and hence broken symmetry between the field theoretic charges and their corresponding partner Hamiltonians Eq.\ (\ref{Q2}). In mechanisms, the source of both of these asymmetries is the absence of bond stretching and hence, forbidden antikink $\mathcal{Q}_2\neq 0$ and forbidden zero mode $A^{\dagger}\psi^{(2)} _0\neq 0$. This is illustrated in Fig.\ (\ref{Schematic}), where we show the symmetry breaking both between a kink and antikink (SUSY field theory), and between the zero modes and the states of self stress around the (anti)kink solution (SUSY QM).

We now define the index $\nu$ of the operator $A$ as the difference in the dimension of the kernel of $A$ and $A^{\dagger}$. Using the identities $\text{ker}A=\text{ker}A^{\dagger} A$ and $\text{ker}A^{\dagger}=\text{ker}A A^{\dagger}$, we obtain the Witten index \cite{Nakahara}
\begin{eqnarray}
\nu=\text{dim ker } {\cal H}_1 - \text{dim ker } {\cal H}_2. 
\end{eqnarray}
The mechanical interpretation of this field-theoretic statement comes from realizing that ${\cal H}_1= A^{\dagger} A$ is a real-space continuum generalization of the discrete dynamical matrix $R^T R$. Hence the dimension of its kernel gives the number of zero-energy displacement modes whereas ${\cal H}_2 = A A^{\dagger}$ (corresponding to $R R^{T} $) gives the states of self-stress. Thus the Witten index (generally defined for supersymmetric theories) reduces to the index obtained in Eq.\ (\ref{eq1}) from the Maxwell count and derived within topological band theory in Ref.\ \cite{kanelubensky}. For the chain mechanism $\nu=1$ -- there is only one normalizable zero-energy eigenstate $\psi^{(1)}_0$ that we associate with $\mathcal{H}_1$.  

Combining ${\cal H}_1,{\cal H}_2$, we get the matrix ${\cal H}$ that together with the operators $\Qm$ and $\Qm^{\dagger}$ given by
\begin{eqnarray}
\mathcal{H}=
\begin{pmatrix}
{\cal H}_1 & 0\\
0 & {\cal H}_2\\
\end{pmatrix}, \quad
\mathcal{Q}=
\begin{pmatrix}
0 & 0\\
A & 0\\
\end{pmatrix}, \quad
\mathcal{Q^{\dagger}}=
\begin{pmatrix}
0 & A^{\dagger}\\
0 & 0\\
\end{pmatrix}
\label{double_hamiltonian}
\end{eqnarray}
satisfies the super-algebra, $[\Hm,\Qm]=[\Hm,\Qm^{\dagger}]=0$, $\{\Qm,\Qm^{\dagger}\}=\Hm$ and $\{\Qm,\Qm\}=\{\Qm^{\dagger},\Qm^{\dagger}\}=0$\cite{witten1,Nakahara}. The two-component field $\Psi_n$, formed  by combining $\psi_n^{(1)}$ and $\psi_n^{(2)}$, can itself be viewed as a ``fermion'' field, as evidenced by the anti-commuting algebra of the $\Qm$ and $\Qm^{\dagger}$ operators.


\section{Mechanical Majorana modes} 
As hinted by the fermionic character of $\Psi$, the same results we have derived in the previous section can be compactly obtained from the Dirac Lagrangian $\mathcal{L}_f$ Eq.\ (\ref{lagrangian}) with inhomogenous mass $w(x)=-\frac{\delta^2W(u)}{\delta u^2}$:
\begin{eqnarray}
\mathcal{L}= i\bar{\Psi}\gamma^{\mu}\partial _{\mu}\Psi + \bar{\Psi}\Psi \!\!\!\! \quad w(x).\label{Dirac}
\end{eqnarray}
The corresponding Euler-Lagrange equation of motion,
\begin{eqnarray}
i\gamma^{\mu}\partial _{\mu}\Psi + w(x) \Psi=0,
\label{Dirac2}
\end{eqnarray}
is a Dirac equation where the constant mass term is replaced by the in-homogenous field $-w(x)$ \cite{JS,JR}. The classical Majorana zero modes minimizes its energy by localizing where $w(x)=-\frac{\delta^2W(u)}{\delta u^2}_{u=u_s(x)}$ is vanishingly small, i.e.\ in the middle of the domain wall for the chain in Fig.\ (\ref{Kink}) or at the core of topological defects in more complex 2D structures \cite{Paulose,JRR}.

We now seek solutions of Eq.\ (\ref{Dirac2}) of the form $\Psi(x,t)=\Psi_n(x)\text{exp}(i\omega_n t)$ and obtain
\begin{eqnarray}
-\gamma^{0}\omega_n \Psi_n(x) + i\gamma^{1}c\frac{d\Psi_n}{dx} + w(x) \Psi_n(x)=0 \!\!\! \quad,
\end{eqnarray}
where the Majorana field $\Psi_n(x)$ and the corresponding gamma matrices $\{\gamma^0,\gamma^1\}$ (Majorana basis) are
\begin{eqnarray}
\Psi_n(x)=
\begin{pmatrix}
\psi^{(1)}_n  \\
\psi^{(2)}_n \\
\end{pmatrix} \quad \gamma^0=
\begin{pmatrix}
0 & -i\\
i & 0\\
\end{pmatrix} \quad
\gamma^1=
\begin{pmatrix}
i & 0\\
0 & -i\\
\end{pmatrix}.
\end{eqnarray}

With the above choices, the two components of the Dirac equation assume the form,
\begin{eqnarray}
A\psi^{(1)} _n(x)=-i\omega_n\psi^{(2)} _n(x),\\
A^{\dagger}\psi^{(2)}_n(x)=i\omega_n\psi^{(1)} _n(x), 
\end{eqnarray}
from which we recover the same eigenvalue problem $A^{\dagger}A\psi^{(1)}_n=\omega_n^2\psi^{(1)}_n$ derived in the previous section. The crucial point is that the operators $A$ and $A^{\dagger}$ are exactly the one derived in Eq.\ (\ref{A_operator}). As a result, $\{\psi_n^{(1)},\psi_n^{(2)}\}$ are the eigenstates of the doubled Hamiltonian ${\cal H}$ in Eq.\ (\ref{double_hamiltonian}). Correspondingly, the zero modes of $A,A^{\dagger}$ are given by Eq.\ (\ref{sech}) and Eq.\ (\ref{zero}) respectively.

\section{Conclusion}

\begin{figure*}
\includegraphics[width=.8\textwidth]{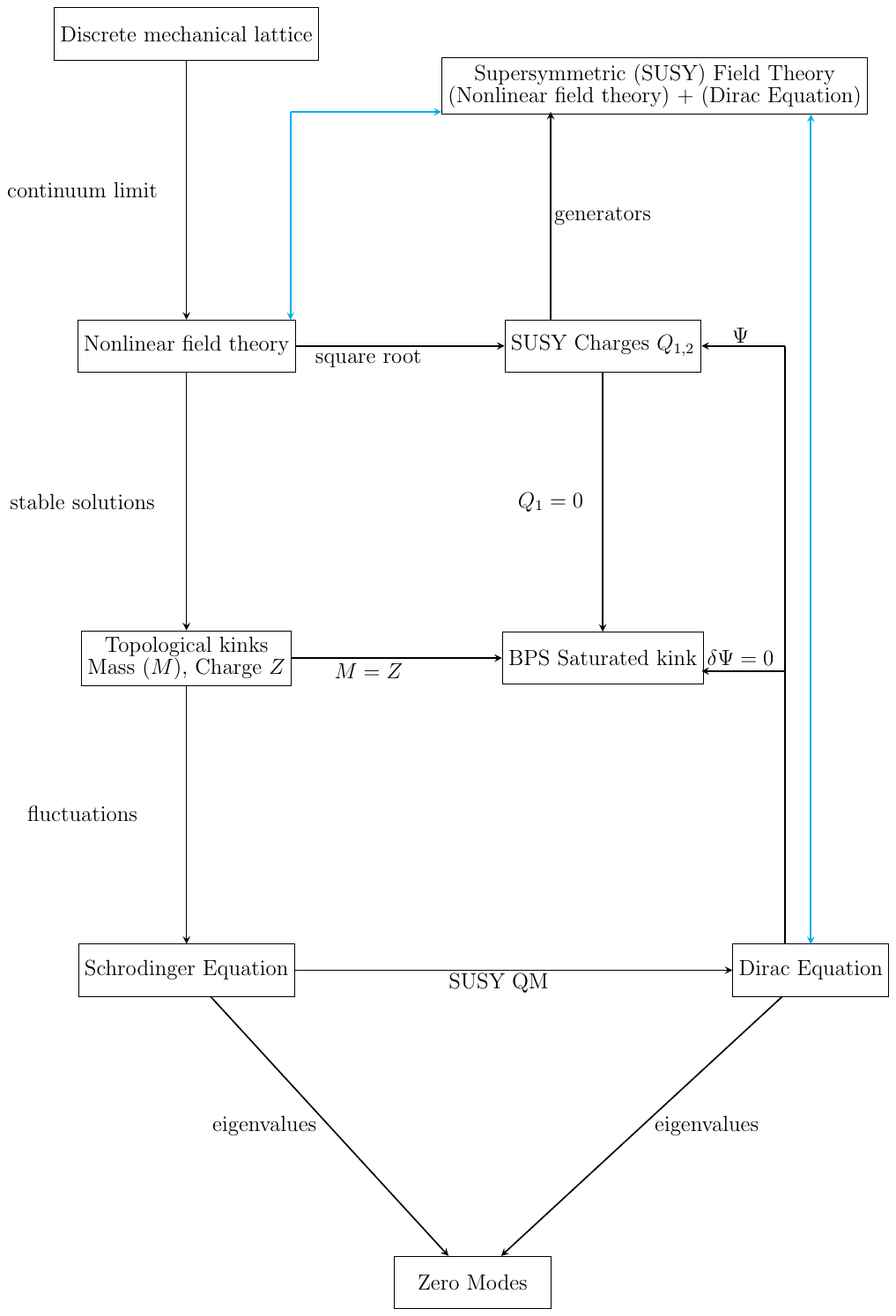} \\
\caption{\label{Map} A map of the theoretical framework used in our work. Beginning with a discrete lattice shown in Fig.\ (\ref{Kink}), we take the continuum limit to obtain Eq.\ (\ref{eqw}) with kink solutions. The study of fluctuations around the kink leads to the Schrodinger equation Eq.\ (\ref{second_order}), which can be decomposed into two first order operators, referred to as charges in supersymmetric quantum mechanics.  The first order operators are in turn identified with the Dirac Equation Eq.\ (\ref{Dirac}). Equivalently, the square root of the nonlinear field theory in Eq.\ (\ref{eqw}) leads to charge Eq.\ (\ref{charge_field}) provided we allow Majorana fermions in our theory. The charge can be identified with conserved charges of the SUSY field theoretic lagrangian Eq.\ (\ref{lagrangian}). }
\end{figure*}

We summarize the theoretical framework and the connections between different ideas used in our work in Fig. (\ref{Map}). Beginning with a discrete lattice shown in Fig.\ (\ref{Kink}), we take the continuum limit to obtain the nonlinear field theory (Eq.\ (\ref{eqw})) with a BPS saturated kink. that comes from enforcing the mechanical constraint Eq.\ (\ref{constraint}). We then generalize Dirac's procedure to take the square root of the nonlinear field theory and obtain fermionic (Majorana) charges, which in turn correspond to symmetries of the Witten-Olive supersymmetric theory Eq.\ (\ref{lagrangian}). 
In order to understand the physical meaning of the fermionic variables, we then study fluctuations around the kink solution and obtain a Schrodinger-like equation Eq.\ (\ref{second_order}), which can be viewed as a product of two first order Dirac-like operators. These Dirac like operators physically describe site displacements and bond extensions of the lattice around stable configurations, which could either be the uniform state (considered in Ref. \cite{kanelubensky}) or the inhomogenous kink state, where nonlinearities are fully taken into account. Further, we show that the Dirac operators can equivalently be obtained as the equation of motion from the fermionic part of the supersymmetric Lagrangian Eq.\ (\ref{lagrangian}). 

Aside from the example explicitly considered here, our approach when run in reverse, could provide a systematic strategy to model non-trivial topological mechanical structures starting from well classified supersymmetric field theories of which the Witten-Olive theory is only the simplest example \cite{Shifman}.

\textbf{Acknowledgments} 
NU and VV were supported by the Complex Dynamics and Systems Program of the Army Research Office under grant no. W911NF-19-1-0268. V.V. acknowledges support from the Simons Foundation. We acknowledge financial support from FOM, NWO and the Delta Institute for Theoretical Physics. We thank A.\ Achucarro, R.\ D.\ Kamien, R.\ J.\ Slager, Y.\ Kafri, E.\ Cobanera, C.\ Beenakker, A.\ Akhmerov, J.\ Paulose, A.\ M.\ Turner and M. Fruchart for stimulating conversations. We thank A. Meeussen for creating movies of the prototypes. We thank ESPCI, Paris for our stay there during the summer of 2014, where part of this work was conceived.

\bibliographystyle{plain}

\begin{thebibliography}{99}
\bibitem{footnote2}The isostatic condition is exactly fulfilled by connecting the two ends of the chain (ring configuration). Once an edge is created, the eigenvalues of the rigidity matrix gives rise to exactly one zero energy mode which will remain localized to an edge of the chain.
\end{thebibliography}

\onecolumngrid

\newpage
\appendix

\section{Supersymmetry generators}\label{App: AppendixA}
We chose our basis such that the $\gamma$ matrices assume the form-
\begin{align}
\gamma^0= \begin{pmatrix}
0  & -i \\
i & 0 \\
\end{pmatrix} \label{Gamma_0}
\end{align}
and
\begin{align}
\gamma^1= \begin{pmatrix}
i  & 0 \\
0 & -i \\
\end{pmatrix}
\end{align}
The supercharges $Q_{1,2}$ are then
\begin{align}
Q_1 &=& \int \ dx \left[\frac{\partial u}{\partial t} \psi _1 + c\left(\frac{\partial u}{\partial x} + \frac{\delta W}{\delta u}\right)\psi _2\right] \\
Q_2&=& \int \ dx \left[\frac{\partial u}{\partial t} \psi _2 + c\left(\frac{\partial u}{\partial x} - \frac{\delta W}{\delta u}\right)\psi _1\right],
\end{align}
The charges generate super-transformations of the fields-
\begin{align}
[Q_{\alpha},u]=-i\psi_{\alpha}, 
\{Q_{\alpha},\bar{\psi}_{\beta}\} = (\gamma^{\nu}\partial_{\nu})_{\alpha\beta}u + i\frac{\partial W}{\partial u}\delta_{\alpha\beta},
\end{align}
where $\alpha,\beta=\{1,2\}$.

Computing $Q^2_1$, we find-
\begin{align}
Q^2_1 = \int\int \ dx \ dx'  \left(\frac{\partial u(x)}{\partial t}\frac{\partial u(x')}{\partial t}\psi _1(x)\psi _1(x')\right) +  c\left(\frac{\partial u(x)}{\partial t}\left[\frac{\partial u(x')}{\partial x'} + \frac{\delta W(u(x'))}{\partial u}\right]\psi _1(x)\psi _2(x')\right) + \nonumber \\ c\left[\frac{\partial u(x)}{\partial x} + \frac{\delta W(u(x))}{\partial u}\right]\frac{\partial u(x')}{\partial t}\psi _2(x)\psi _1(x') + c^2\left[\frac{\partial u(x)}{\partial x} + \frac{\delta W(u(x))}{\partial u}\right]\left[\frac{\partial u(x')}{\partial x'} + \frac{\delta W(u(x'))}{\partial u}\right]\psi _2(x)\psi _2(x'). 
\label{Q12}
\end{align}
By interchanging the order of integrals involving variables $x,x'$, we can express Eq.\ (\ref{Q12}) in the form-
\begin{align}
2Q^2_1 = \int\int \ dx \ dx'  \left(\frac{\partial u(x)}{\partial t}\frac{\partial u(x')}{\partial t}\{\psi _1(x),\psi _1(x')\}\right) + c\left(\frac{\partial u(x)}{\partial t}\left[\frac{\partial u(x')}{\partial x'} + \frac{\delta W(u(x'))}{\partial u}\right]\{\psi _1(x),\psi _2(x')\}\right) +\nonumber  \\ c\left[\frac{\partial u(x)}{\partial x} + \frac{\delta W(u(x))}{\partial u}\right]\frac{\partial u(x')}{\partial t}\{\psi _2(x),\psi _1(x')\} +  c^2\left[\frac{\partial u(x)}{\partial x} + \frac{\delta W(u(x))}{\partial u}\right]\left[\frac{\partial u(x')}{\partial x'} + \frac{\delta W(u(x'))}{\partial u}\right]\{\psi _2(x),\psi _2(x')\}. \label{Q1square}
\end{align}

We next make use of the anti-commutation relation \cite{Shifman}
\begin{align}
\left\{\psi _{\alpha}(x,t),\bar{\psi} _{\beta}(x',t)\right\}=\left(\gamma^0\right)_{\alpha\beta}\delta(x-x'), \label{anti}
\end{align}
where, $\bar{\psi}=\psi^{\dagger}\gamma ^0$. Therefore, $\bar{\psi}_1 = i\psi _2$ and $\bar{\psi}_2=-i\psi _1$. Thus,
 \begin{align}
 \left\{\psi _{1}(x,t),\psi _{1}(x',t)\right\}=  i\left\{\psi _{1}(x,t),\bar{\psi} _{2}(x',t)\right\}= \delta(x-x') , \nonumber \\
 \left\{\psi _{2}(x,t),\psi _{2}(x',t)\right\}=  -i\left\{\psi _{2}(x,t),\bar{\psi} _{1}(x',t)\right\}= \delta(x-x') , \label{anti2}
 \end{align} 
with the rest being 0. 

Substituting Eqs.\ (\ref{anti2}) into Eq.\ (\ref{Q1square}), we obtain Eq.\ (\ref{Q2}) in the main text. 

\newpage
\section{Rigidity matrix}\label{App: AppendixB}

The Lagrangian for a chain with $n$-rotors is
\begin{align*}
\mathcal{L} = \sum _n \frac{1}{2}I\left(\frac{d\theta _n}{dt}\right)^2 - \frac{1}{2}K\left(l_{n,n+1}-\bar{l} \right)^2,
\end{align*}
where, $I=mr^2$ is the moment of Inertia of a rotor of length $r$ and mass $m$. $K$ is the bare spring constant of the bar joining two adjacent rotors at sites $n,n+1$ whose squared length can be given by the expression \cite{kanelubensky}
\begin{align}
l^2_{n,n+1}=a^2 + 2r^2\cos(\theta _n+\theta _{n+1})+2ar\left[\sin\theta_n-\sin\theta _{n+1}\right] \label{length}
\end{align}
where, $\theta$ is the angle with respect to the vertical measured positive in the clockwise direction and $a$ is the lattice spacing. In the uniform state, $|\theta _n|=|\theta _{n+1}|=\bar{\theta}$ and $l_{n,n+1}=\bar{l}=\sqrt{a^2+4r^2\cos^2\bar{\theta}}$ for all $n$. The Euler-Lagrange equation of motion for the $n-$th rotor is 
\begin{align}
I\ddot{\theta} _{n} = -K(l_{n,n+1}-\bar{l})\frac{\partial l_{n,n+1}}{\partial \theta _n}-K(l_{n-1,n}-\bar{l})\frac{\partial l_{n-1,n}}{\partial \theta _n} \label{EOM1}.
\end{align}
Shifting the index $n$ by one unit,  we obtain:
\begin{align}
I\ddot{\theta} _{n-1} = -K(l_{n-1,n}-\bar{l})\frac{\partial l_{n-1,n}}{\partial \theta _{n-1}}-K(l_{n-2,n-1}-\bar{l})\frac{\partial l_{n-2,n-1}}{\partial \theta _{n-1}} \label{EOM2}.
\end{align}

Within linear approximation in the angular displacements (i.e., $\theta _n=\bar{\theta} - \delta _n$), Eq.\ (\ref{length}) can be approximated as
\begin{align*}
l^2_{n,n+1} = \bar{l}^2  + 2r^2\sin 2\bar{\theta}(\delta _n+\delta _{n-1}) -2ar\cos\bar{\theta}(\delta _n-\delta _{n+1}).
\end{align*}
from where, the infinitesimal change in the length of the spring $l^2_{n,n+1}=(\bar{l}+\delta l_{n,n+1})^2\approx \bar{l}^2 + 2\bar{l}\delta l_{n,n+1}$  can be expressed in the form  
\begin{align}
\delta l_{n,n+1} = q_+\delta _n + q_-\delta _{n+1},  
\end{align} 
where, $q_{\pm}=\frac{r\cos\bar{\theta}\left(2r\sin\bar{\theta}\pm a\right)}{\bar{l}}$.

The linearized equation of motion Eq.\ (\ref{EOM1}-\ref{EOM2}) is then 
\begin{eqnarray}
-\ddot{\delta} _n &=& (q^2_+ + q^2_-)\delta _n + q_+q_-(\delta _{n+1}+\delta _{n-1}),\nonumber \\
-\ddot{\delta} _{n-1} &=& (q^2_+ + q^2_-)\delta _{n-1} + q_+q_-(\delta _{n}+\delta _{n-2}). 
\label{Linear_EOM}
\end{eqnarray}

Expressing Eqs.\ (\ref{Linear_EOM}) in terms of normal coordinates with the choice  $\delta _{n-2}=c_1e^{i\left(k(n-2)a-\omega t\right)}$, $\delta _{n-1}=c_2e^{i\left(kna-\omega t\right)}$,$\delta _n=c_1e^{i\left(kna-\omega t\right)}$, $\delta _{n+1}=c_2e^{i\left(k(n+2)a-\omega t\right)}$, where $k$ is the wavenumber, we obtain a matrix equation:
\begin{align*}
\omega^2\begin{pmatrix}
c_1 \\
c_2 \\
\end{pmatrix}=
D(k)\begin{pmatrix}
c_1 \\
c_2 \\
\end{pmatrix}
\end{align*}
where, $D(k)$ is the two by two dynamical matrix \cite{kanelubensky,Bryan}
\begin{align}
D(k) = \begin{pmatrix}
q^2_++q^2_- & (1+e^{2ika})q_+q_- \\
(1+e^{-2ika})q_+q_- & q^2_++q^2_- \\
\end{pmatrix}.
\end{align}
\\\\
In order to take the square root of the dynamical matrix, we seek a matrix $R(k)$, such that $R^{\dagger}(k)R(k)={D}(k)$ \cite{kanelubensky}. Consider a general 2 by 2 complex matrix
\begin{align}
R(k) = \begin{pmatrix}
w  & x \\
y & z \\
\end{pmatrix}
\end{align}
where, $w,x,y,z$ are as yet undetermined complex functions of $k$. In order to find these  functions, we first evaluate $R^{\dagger}(k)$ and then compare the product of $R^{\dagger}R$ to $D(k)$. This gives us the following three relations-
\begin{align}
|w|^2+|y|^2 &=& q^2_+ +q^2_- \label{Constraint1}\\
|x|^2+|z|^2  &=& q^2_+ +q^2_-\label{Constraint2}\\
w^*x + y^*z &=& q_+q_-(1+e^{2ika}). \label{Constraint3}
\end{align}

A possible choice that satisfies the first constraint Eq.\ (\ref{Constraint1}) is  $w=q_+$ and $y=q_-$. Next, if we chose $x=q_-e^{i0}$ and $z=q_+e^{2ika}$, we can satisfy both the constraint equations Eqs.\ (\ref{Constraint2}-\ref{Constraint3}). Thus, a possible choice for $R(k)$ is-
\begin{align}
R(k) = \begin{pmatrix}
q_+ & q_- \\
q_- & q_+e^{2ika} \\
\end{pmatrix}.
\end{align}
This is the Rigidity matrix in Fourier space. 

Physically,the rigidity matrix relates site displacements $u$ to bond elongations $\delta l$ in real space. To identify the first order differential operator $A$ in Eq. (\ref{A_operator}) with the rigidity matrix in real space, consider again a pair of adjacent sites $n,n+1$ in Fig.\ (1). The bond extensions are $\delta l_{n,n+1} = \frac{2}{\bar{l}}(\frac{a}{2}r\cos\bar{\theta}(\delta\theta _n- \delta\theta _{n+1}) + r^2\sin \bar{\theta}\cos\bar{\theta}(\delta\theta_n+\delta\theta _{n+1}))$ where, $\delta\theta _n,\delta\theta _{n+1}$ are small angular displacements around the {\it homogeneous} background $\theta=\bar{\theta}$. 

A continuum limit of the distortion field $r\cos\bar{\theta}\delta\theta _n \rightarrow u(x),r\cos\bar{\theta}(\delta\theta_n-\delta\theta _{n+1})\rightarrow -\partial _xu$, reproduces the operator $A$ in Eq.\ (\ref{A_operator}) for the special case of a constant potential $w(x)= -\left(\frac{dV}{du}\right)_{u=\bar{u}}=2r\sin\bar{\theta}=2\bar{u}$. However, when we expand around the soliton field as in Eq.\ (\ref{second_order}), we obtain bond extensions over an inhomogenous zero-energy state. Thus in general, the operator $A$ is a continuum limit of the real-space rigidity matrix around a specific solution of the non-linear field theory and can be explicitly determined using Eqs.\ (\ref{A_operator}) and the superpotential $w(x)$. 


\end{document}